\documentclass[3p,times]{elsarticle}

\usepackage{ecrc}

\usepackage{epstopdf}

\volume{00}

\firstpage{1}

\journalname{Nuclear Physics A}

\runauth{Author1 et al.}

\jid{nupha}

\jnltitlelogo{Nuclear Physics A}

\usepackage{graphicx}
\usepackage{amsmath,amssymb}

\begin{document}

\begin{frontmatter}



\title{Quantitative extraction of the jet transport parameter from combined data at RHIC and LHC}

\author{Xin-Nian Wang (for the JET Collaboration)}
\address{Key Laboratory of Quark and Lepton Physics (MOE) and Institute of Particle Physics,
Central China Normal University, Wuhan 430079, China}
\address{Nuclear Science Division Mailstop 70R0319, Lawrence Berkeley National Laboratory, Berkeley, California 94740, USA}




\begin{abstract}
Using theoretical tools developed by the JET Collaboration in which one employes 2+1D or 3+1D hydrodynamic models for the bulk medium evolution and jet quenching models, the combined data on suppression of single inclusive hadron spectra at both RHIC and LHC are systematically analyzed with five different approaches to the parton energy loss. The jet transport parameter is extracted from the best fits to the data with values of $\hat q \approx 1.2\pm 0.3$ and $1.9\pm 0.7$ GeV$^2$/fm in the center of the most central Au+Au collisions at $\sqrt{s}=200$ GeV and Pb+Pb collisions at $\sqrt{s}=2.67$ TeV, respectively at an initial time $\tau_0=0.6$ fm/$c$ for a quark jet with an initial energy of 10 GeV/$c$. 
\end{abstract}


\end{frontmatter}



\section*{}

 Jet quenching was originally proposed as one of the signatures of the formation of hot QCD matter in high-energy heavy-ion collisions \cite{Wang:1991xy}. The observation of the jet quenching phenomenon \cite{Adcox:2001jp} more than 10 years ago in the first RHIC experimental results marked a historical milestone that the jet quenching had gone from a theoretical concept to experimental studies in reality. Since then there have been a plethora of jet quenching phenomena in high-energy heavy-ion experiments, from suppression of single hadron spectra to suppression of back-to-back dihadron and gamma-hadron correlation at both RHIC and LHC. The most direct observation of jet quenching at LHC is through mono-jet events in ATLAS and CMS experiments at LHC \cite{Muller:2012zq}. We have come currently to  a cross point at which we have to move the discussion about jet quenching from pure observation to using it as a tool to study properties of the dense medium in heavy-ion collisions. Among the fundamental properties of the quark-gluon plasmas, we are interested in the space-time profile of the local temperature, fluid velocity, the equation of state, sound velocity, EM correlation and bulk medium transport coefficients. The ultimate goal of future studies is to measure or extract these medium properties from comparisons between experimental data and theoretical calculations.  Jet transport parameter, defined as the averaged transverse momentum broadening squared per unit length which is also related to the local gluon number density,
 \begin{equation}
 \hat q= \frac{4\pi^2\alpha_s C_R}{N_c^2-1}\int\frac{dy^-}{\pi}\langle F^{\sigma +}(0) F_{\sigma}^{\,+}(y)\rangle=\frac{4\pi^2\alpha_s C_R}{N_c^2-1}\rho_A xG_N(x)|_{x\rightarrow 0},
 \end{equation}
 is one of the medium properties that one should be able to extract from experimental data on jet quenching.
 
Jets in high-energy collisions are produced through hard processes that can be calculated with perturbative QCD. Single jet production in deeply inelastic lepton-nucleus scattering (DIS) is the simplest example of such hard processes. In this case, the virtual photon kicks a quark off a nuclear target that will eventually hadronize into the final jet of hadrons. To calculate the transverse momentum distribution of the struck quark, one has to consider its interaction with the remnant target through soft and collinear gluons. Inclusion of such multiple gluon interactions is also important to ensure the gauge invariance of quark transverse momentum distribution. Through Taylor expansion of the Fourier transform of transverse-momentum-dependent (TMD) quark distribution in transverse coordinate, one can express the TMD distribution as a power series in a collinear operator, referred to as the jet transport operator \cite{Liang:2008vz}. If one knows values of the matrix elements of these jet transport operators, one should know the quark TMD distribution. Assuming two-gluon correlation approximation or equivalently random walk in transverse diffusion through color Lorentz force, the transverse momentum distribution of the final quark can be cast into a Gaussian form with the width given by the path length integral of the quark transport parameter $\hat q$. Such transverse momentum broadening has been measured in DIS experiments through transverse momentum distribution of leading hadrons and suppression of leading hadron spectra. The extracted value of jet transport parameter in large cold nuclei is $\hat q_N\approx 0.02$ GeV$^2$/fm \cite{Deng:2009qb}.
 
In heavy-ion collisions, the physics picture is similar but with much more complicated medium that is no longer static as in a cold nucleus. The hot QCD medium is transient with a very short life-time and rapid expansion. One has to take into account the dynamical evolution of the bulk medium to accurately describe jet quenching phenomena. A realistic hydrodynamical model becomes a necessary tool not only for a quantitative description of the spectra of soft bulk hadrons but also for jet quenching that governs the jet or hadron spectra at high $p_T$. One therefore needs a comprehensive framework for the study of both soft hadrons and jet quenching and extraction of bulk transport coefficients and jet transport parameter. This is the purpose of the JET topical collaboration. The final goal of the JET Collaboration is to create a comprehensive Monte Carlo package which combines the most advanced and experimentally constrained model for bulk medium evolution, up-to-date models for parton propagation and transport in medium and final hadronization of jet shower partons and jet-induced medium excitation. Since the start of the JET Collaboration more than 3 years ago, significant progresses have been achieved towards this final goal.  An enormous database of spatial-time profiles of bulk medium evolution in high-energy heavy-ion collisions have been generated from event-by-event viscous hydrodynamic model calculations and stored in a server for any public use \cite{jetbulk}. As an intermediate step toward the final goal of the JET Collaboration, a jet package has been created that employs semi-analytic calculations through Monte Carlo integration for single hadron spectra in heavy-ion collisions, which will also include dihadron and gamma-hadron correlation in the future. Five different approaches to parton energy loss have been implemented in this jet package together with bulk medium evolution from 2+1D and 3+1D hydrodynamic models that are constrained by the bulk hadron spectra and their anisotropic flows. These five approaches to parton energy loss include: GLV \cite{Gyulassy:2000er} and its recent CUJET implementation \cite{Buzzatti:2011vt} which uses a potential model for multiple parton scattering in the QGP  medium; the high-twist (HT) approaches (HT-BW and HT-M) \cite{Chen:2011vt,Majumder:2011uk} have the jet transport parameter as the only medium property that enters parton energy loss; The MARTINI \cite{Schenke:2009gb} and McGill-AMY \cite{Qin:2007rn} model are based on hard-thermal-loop (HTL) resummed thermal field theory. In GLV, MARTINI and Mc-Gill-AMY, the only adjustable parameter is the strong coupling constant with which one can in turn calculate the jet transport parameter $\hat{q}$ for a quark jet. 
 
\begin{figure}[h]
\centering
\includegraphics[width=8cm]{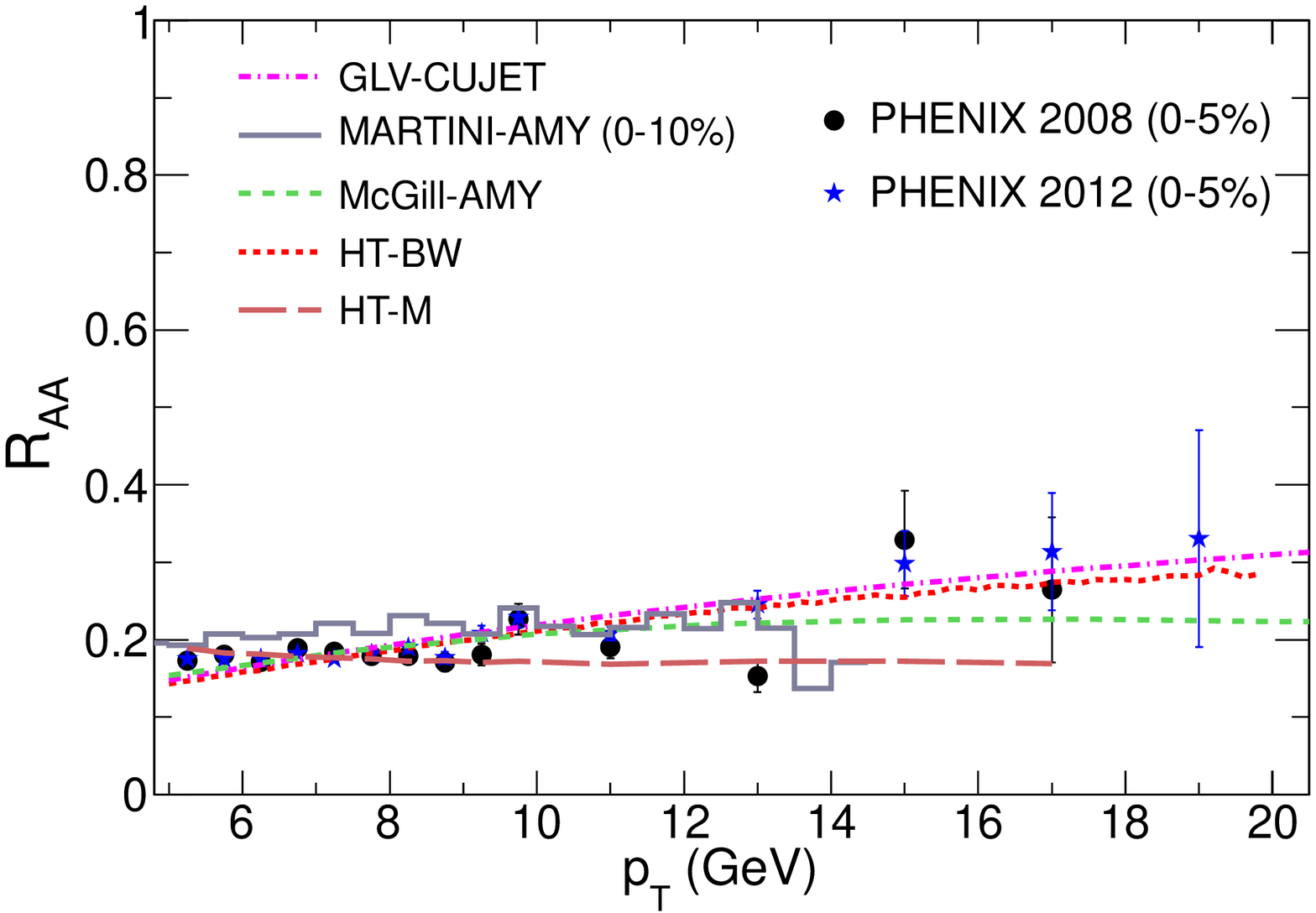} \includegraphics[width=8cm]{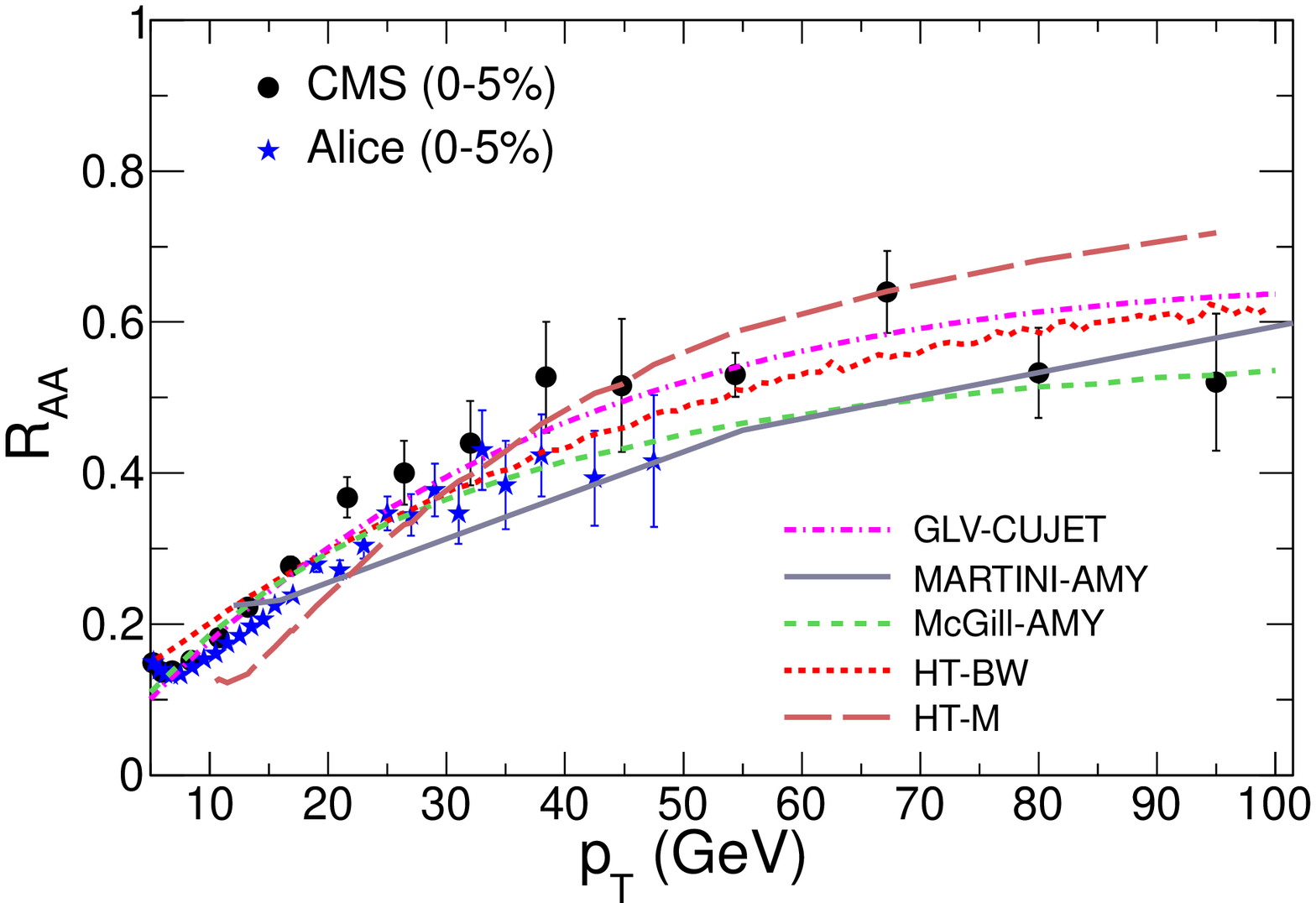}
\caption{(Color online) Best fits to the suppression ratios of single hadron spectra in central (0-5\%) Au+Au at $\sqrt{s}=200$ GeV (left panel) and Pb+Pb collisions at $\sqrt{s}=2.76$ TeV (right panel) as compared to PHENIX data \cite{Adare:2008qa,Adare:2012wg} at RHIC and ALICE \cite{Abelev:2012hxa} and CMS data \cite{CMS:2012aa} at LHC.} 
\label{fig1}
\end{figure}

 Using this jet package, we have carried out a comprehensive analysis of experimental data on suppression of single hadron spectra in heavy-ion collisions at both RHIC and LHC \cite{Burke:2013yra}. Within each approach of parton energy loss, a $\chi^2$ fit to the experimental data produces a set of model parameters that correspond to given values of the jet transport parameter.  Shown in Fig.~\ref{fig1} are the best fits to single hadron spectra suppression ratios in the most central (0-5\%) Au+Au collisions at RHIC and Pb+Pb collisions at LHC with five different approaches to parton energy loss. Experimental data at LHC and their strong $p_T$ dependence are very critical to constrain different numerical implementations of parton energy loss and have ruled out some earlier models that gave unrealistic large values of $\hat q$. LHC data have also forced many improvements to some of the five models shown in Fig.~\ref{fig1}, for example, running coupling constant in CUJET and MARTINI. 
 
With these fits to the combined experimental data at RHIC and LHC, one can extract values of the jet transport parameter $\hat q$ at the center of the most central heavy-ion collisions at an initial time $\tau_0=0.6$ fm/$c$ for a quark jet with an initial energy of 10 GeV.  For the hydrodynamic models used in this study, the corresponding initial temperatures are $T_0=346-373$ and 447-486 MeV at the center of the most central Au+Au collisions at $\sqrt{s}=200$ GeV/n at RHIC and Pb+Pb collisions at $\sqrt{s}=2.76$ TeV/n at LHC, respectively.

\begin{figure}
\centering
\includegraphics[width=8cm]{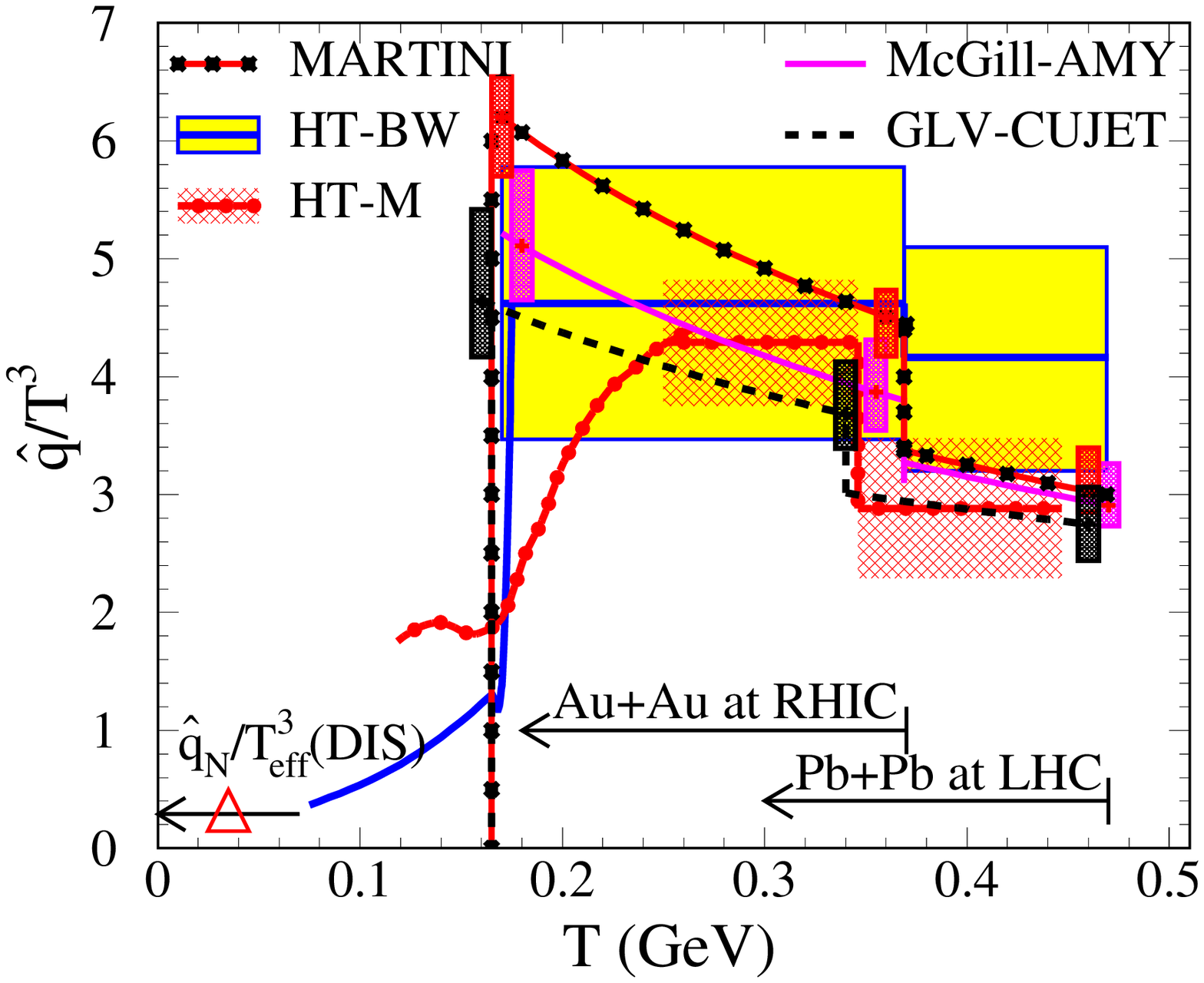} \includegraphics[width=8cm]{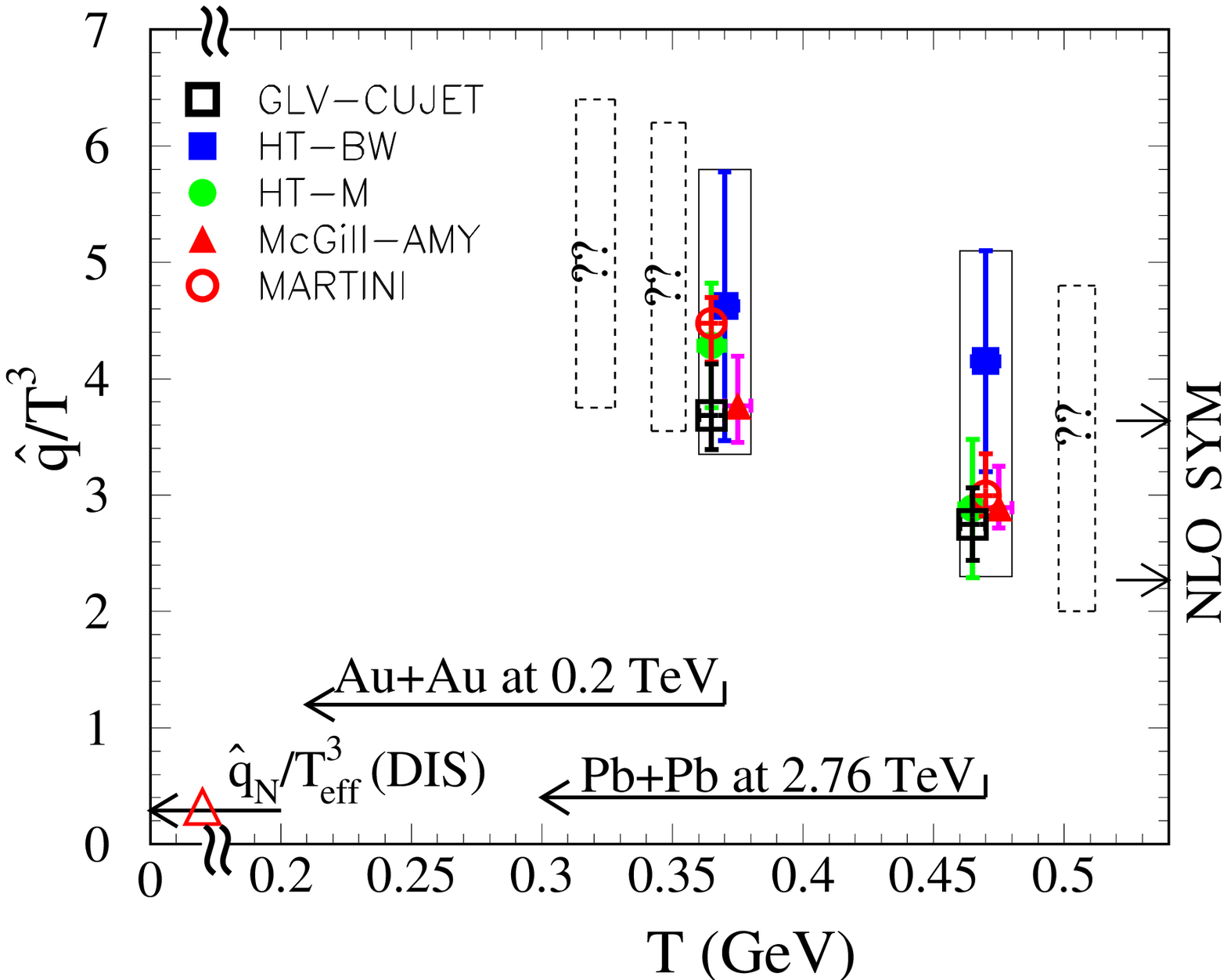}
\caption{\label{fig2} (Color online) (right panel) The extracted jet transport parameter $\hat q/T^3$ with different approaches to jet quenching for an initial quark jet with energy $E=10$ GeV at the center of the most central A+A collisions at an initial time $\tau_0=0.6$ fm/$c$. Errors from the fits are indicated by filled boxes at three separate temperatures at RHIC and LHC, respectively. The arrows indicate the range of temperatures at the center of the most central A+A collisions. The triangle indicates the value of $\hat q_N/T^3_{\rm eff}$ in cold nuclei from DIS experiments. Lines are the assumed temperature dependence of $\hat q$ in each model. (left panel) The dashed boxes indicate expected values in A+A collisions at $\sqrt{s}=0.063, 0.130$ and 5.5 TeV/n. Values of $\hat q_{\rm SYM}^{\rm NLO}/T^3$ from NLO SYM theory are indicated by two arrows on the right axis.} 
\label{fig2}
\end{figure}

 Shown in Fig.~\ref{fig2} (right panel) are the extracted or calculated values of $\hat q$ as a function of the initial temperature. Both HT approaches assume non-vanishing values of $\hat q$ in hadronic phase. In CUJET, MARTINI and McGill-AMY $\hat q$ is calculated within each model, which should have a logarithmic dependence in both energy and temperature, with the extracted values of strong coupling constant $\alpha_{\rm s}$ . The HT models assume that $\hat q$ is independent of jet energy in this study. The errors within each model are from the $\chi^2$ fit with one standard deviation. The variation of $\hat q$ values between different models can be considered as theoretical uncertainties. One therefore can extract its range of values at RHIC and LHC as:
\begin{equation*}
\frac{\hat q}{T^3}\approx \left\{ 
\begin{array}{l}
4.6\pm 1.2 \qquad \text{at RHIC},\\
3.7 \pm 1.4 \qquad \text{at LHC},
\end{array}
\right.
\end{equation*}
at the highest temperatures reached in the most central Au+Au collisions at RHIC and Pb+Pb collisions at LHC. 
The corresponding absolute values for $\hat q$ for a 10 GeV quark jet are,
\begin{equation*}
\hat q \approx \left\{ 
\begin{array}{l}
1.2 \pm 0.3  \\
1.9 \pm 0.7 
\end{array}
 \;\; {\rm GeV}^2/{\rm fm} \;\; \text{at} \;\;
\begin{array}{l}
 \text{T=370 MeV},\\
 \text{T=470 MeV},
\end{array}
\right.
\end{equation*}
at an initial time $\tau_0=0.6$ fm/$c$. These values are consistent with LO pQCD estimates, however, with a somewhat small values of $\alpha_{\rm s}$ as obtained in CUJET, MARTINI and McGill-AMY model.  The value of $\hat q_N /T^3_{\rm eft}$ in cold nuclei as extracted from jet 
quenching in DIS \cite{Deng:2009qb} is also shown here.  The value of $\hat q_N=0.02$ GeV$^2$/fm is an order of magnitude smaller than that in A+A collisions at RHIC and LHC.

In the immediate future, one should be able to carry out the same analyses at higher LHC energy and some of the beam scanning energies at RHIC. Shown in Fig.~\ref{fig2} (left panel) as open boxes with question marks are the predicted values of $\hat q$ at future higher LHC energy and RHIC bean scanning energies. Together with the current values at the LHC and the highest RHIC energy, one can obtain a rough temperature dependence of $\hat q/T^3$. Furthermore, comparisons to dihadron and gamma-hadron correlations can provide additional constraints on $\hat q$. 

In the long term future, one should develop and implement complete next-to-leading order calculations of parton energy loss for further reduction of theoretical uncertainties. Though factorization of initial jet production and final-state parton energy loss is assumed in all studies, it has never been explicitly proven nor illustrated. In a recent study, the complete NLO calculations of transverse 
 momentum weighted cross sections of semi-inclusive DIS (SIDIS) and Drell-Yan processes in p+A collisions have been performed for the first time at twist-four \cite{Kang:2013raa}. The factorization of the initial production hard processes and higher-twist matrix elements or $\hat q$ in the final-state has been explicitly illustrated. Furthermore, the QCD evolution of the twist-four parton correlation matrix which is related to $\hat q$ has been identified. One therefore can solve the evolution equation and determine the scale dependence of the jet transport parameter $\hat q$. Such evaluation of $\hat q$ evolution has also been similarly performed \cite{Iancu}. This should be one of the long-term goals of experimental studies of jet quenching in future high-energy heavy-ion collisions.
 
 As a final remark, I would like to emphasize that $\hat q$ represents the averaged transverse momentum broadening squared for single partons. Therefore, broadening of dijet correlation in azimuthal angle is only indirectly related to $\hat q$. In a recent study within Linear Boltzmann Transport (LTB) model, the large angle tail of the dijet correlation is shown to be sensitive to the value of $\hat q$ \cite{Wang:2013cia}. High precision data are needed for any phenomenological study. The most sensitive probe of jet quenching and $\hat q$ is still gamma-hadron correlation in azimuthal angle which is also a sensitive probe of jet-induced medium excitations \cite{Li:2010ts,Ma:2010dv}.
 
 This work is supported by China MOST under Grant No. 2014DFG02050, the NSFC under Grant No. 11221504, the Major State Basic Research Development Program in China (No. 2014CB845404), U.S. DOE under Contract No. DE-AC02-05CH11231 and within the framework of the JET Collaboration.













\end{document}